# A metrological investigation of (6,5) carbon nanotube absorption cross-section

*Laura Oudjedi[1,2], A. Nicholas G. Parra-Vasquez[1,2,§], Antoine G. Godin[1,2], Laurent Cognet[1,2], and Brahim Lounis[1,2,*]*

[1] Univ Bordeaux, LP2N, F-33405 Talence, France

[2] CNRS & Institut d'Optique, LP2N, F-33405 Talence, France

Abstract

Using single nanotube absorption microscopy, we measured the absorption cross-section of (6,5) carbon nanotubes at their second order optical transition. We obtained a value of 3.2 $10^{-17}$ cm$^2$ per carbon atom with a precision of 15% and an accuracy below 20%. This constitutes the first metrological investigation of the absorption cross-section of chirality-identified nanotubes. Correlative absorption-luminescence microscopies performed on long nanotubes reveal a direct manifestation of exciton diffusion in the nanotube.

Single wall carbon nanotubes (SWCNTs) provide unique opportunities for applications in electronics, optoelectronics, photonics and photovoltaics[1]. They are indeed nearly ideal models of infinite π orbital conjugations where the *sp²* lattice provides exceptional charge carrier properties[2]. Moreover, their true one-dimensional character confers them well-defined optical transitions from excitonic states[3, 4, 5]. While their Raman and photoluminescence properties have been extensively studied[5], the basic absorption properties of SWCNT are still poorly established in a quantitative manner, even if they are of prime importance for applications. Indeed, common absorption measurements on ensemble of SWCNT suffer from the problem of samples heterogeneities inherent to standard synthesis methods. The large multiplicity of nanotube chiralities and lengths, as well as the presence of synthesis impurities impede to assess the exact amount of a SWCNT chirality contributing to an absorption signature. For (6,5) nanotubes, absorption cross-sections $\sigma_C$ ranging from $0.5 \times 10^{-18}$ to $1.3 \times 10^{-17}$ cm² per C atom can be deduced from previous reports[6-9] considering an excitation light polarized along the nanotubes and resonant with the second order transition ($S_{22}$). The majority of the caveats mentioned above which are responsible for the disparity in $\sigma_C$ determination can be lifted-off by performing measurements on individual SWCNTs.

Single SWCNT absorption measurements are challenging because the weak signals have to be extracted from laser intensity fluctuations (including the always present shot noise) and contributions from environment scattering[10]. Luminescence microscopy, which is an intrinsic dark field method, has been used to perform the first evaluation of $\sigma_C$ for individual (6,5) nanotubes. The method was based on the study of the luminescence dependence with excitation intensity, which deviates from linearity at high excitation due to exciton-exciton annihilation processes[11]. From the saturation intensity and the luminescence lifetime of the SWCNT $\sigma_C$ was estimated to be $\sim 10^{-17}$ cm² per C atom[11]. However, individual defects that are photoinduced upon increasing excitation intensity quench the luminescence and therefore can strongly alter the saturation behavior[12, 13]. Rayleigh intensity scattering technique combined with AFM have been used to extract an estimate of $\sim 2.5 \times 10^{-17}$ cm²/C for the resonant absorption cross section of nanotubes of different diameters[14]. Recently, a value in the range of $\sim 10^{-17}$ cm²/C was reported for a large diameter (18,5) nanotube lying on opaque substrate excited around its $S_{33}$ transition using a reflective modulation imaging technique[15].

Given the large disparity in the reported absorption cross-section values, a metrological approach is essential to provide a definitive value. Here, we present a quantitative measurement of the absorption cross-section of chirality identified individual (6,5) nanotubes at their $S_{22}$ transition peak. We choose the (6,5) chirality as standard nanotubes since they can easily be identified by their luminescence spectra and be excited efficiently near their second order resonance $S_{22}$ using low noise CW solid-state lasers. Futhermore, they are widely studied because of their abundance in standard synthesis methods (CoMoCAT or HiPco for instance) and because their luminescence falls in the detection window of

Silicon detectors. Our direct measurement of carbon nanotube absorption cross-section requires neither assumption on the sample content, nor on data interpretation. It was made possible by combining transmission microscopy for absorption determination with luminescence microscopy for chirality assignment and photothermal microscopy for assuring nanotube isolation.

We optimize the transmission spatial modulation technique[16] to detect the extinction cross section of small diameter SWNTs with high signal to noise ratios. For small diameter SWCNTs, the extinction cross section is dominated by the absorption as demonstrated by Rayleigh scattering experiments which reported scattering cross-section of the order of a few $10^{-20}$ cm²/C atom for (6,4) nanotubes[14]. One limiting factor of this technique was that only relatively low frequencies (< kHz) could be used for modulating the sample position due to mass loading[17]. Thus, these measurements suffer from 1/f noise. Our approach is based on a high frequency ($f$ = 100 kHz) spatial modulation of a tightly focused Gaussian beam across the nanoobject (Figure 1). In practice, the spatial modulation of a 561nm laser beam is performed by varying the first-order diffraction angle of an acousto-optic modulator. The incident beam is focused onto the sample by means of a high NA microscope objective (x60, NA 1.45) and its transmission collected by a second identical objective is sent to a balanced photodetector connected to a lock-in amplifier for demodulation. Images of the samples are obtained by raster scanning the sample mounted on a piezo-scanning stage.

The sensitivity and metrological potential of the method is first validated on a standard nanosphere sample. This sample consists of gold nanoparticles (NP) with mean diameter 10.0±0.8 nm (measured on 191 NPs by TEM, figure 2 a-b) deposited on cleaned glass coverslip, embedded in immersion oil for index matching (~1.50). In the case of small nano-objects the extinction cross section is dominated by the absorption[18] (negligible intensity scattering). For a small spherical gold NP (with dimensions much smaller than the wavelength) at the position (x, y), the transmitted beam power writes $P_{trans}(t) = P_{inc} - \sigma_{Au} I(x, y + \delta \sin(2\pi f t))$ with $P_{inc}$ the incident beam power, $\sigma_{Au}$ the NP absorption cross-section, $I(x,y)$ the intensity profile at the sample, and $\delta$ the amplitude of the spatial modulation along the y-axis. Assuming a Gaussian beam profile $I(x',y') = I_0 e^{\frac{-2(x'^2+y'^2)}{w_0^2}}$ and a small beam modulation amplitude ($\delta \ll w_0$) the demodulated power is proportional to the beam profile first derivative along the y-axis and writes $P_{mod}(x,y) = -\frac{8y}{\pi w_0^4} \sigma_{Au} \delta \times P_{inc} e^{\frac{-2(x^2+y^2)}{w_0^2}}$. As shown on figure 2c-d, individual gold nanoparticles are imaged with a high signal-to-noise-ratio (>20) with short integration times of 25 ms and excitation power of 650 µW.

It follows that $\sigma_{Au}$ can be directly measured from the peak-to-peak amplitude $A_{pp}$ of the profile ( $\sigma_{Au} = \frac{\pi e^{1/2}}{8} \frac{w_0^3}{\delta} \frac{A_{pp}}{P_{inc}}$ ) given the precise determination of the beam size $w_0$ and displacement $\delta$.

For this purpose, the beam profile was determined by confocal fluorescence images of 45 individual 20 nm luminescent spheres deposited on a glass coverslip (Figure 1S, Supp. Info). We find $w_0 = 269 \pm 6$ nm (mean ± standard deviation). The beam displacement at the sample is calibrated by measuring the displacement of individual quantum dots positions found in two confocal images acquired with two different RF waves driving the acousto-optic modulator (Figure 2S, Supp. Info). From 13 acquisitions, we found $\delta = 36.4 \pm 1.9$ nm (mean ± standard deviation).

After careful calibration of the demodulated signal, one obtain a mean absorption cross-section for 10 nm gold NPs of 41±16 nm$^2$ (mean ± half width at half maximum, N=191) at 561 nm excitation wavelength (Figure 2e). This is in excellent agreement with the value of 42 nm$^2$ predicted by the Mie theory[18, 19] for a particle surrounded by a medium of refraction index of 1.50. Noteworthy, the dispersion in the measured NP absorption cross-section (33%) is mainly imposed by the NP size dispersion (30% in volume), since the measurement uncertainties play here a minor role (see below). This measurement demonstrates that our experimental procedure allows quantitative determination of small nano-object absorption cross-sections.

We then consider (6,5) SWCNTs to achieve precise characterization of their absorption cross-section. Raw HiPco SWCNTs were solubilized in 1% aqueous sodium deoxycholate (DOC) using brief tip sonication followed by bench-top centrifugation to remove un-solubilized material from the suspension. Nanotubes are deposited on cleaned glass coverslip and covered with immersion oil for further analysis. In order to identify long individual (6,5) SWCNTs, single-molecule wide-field photoluminescence (PL) microscopy is first performed with a detection window around their $S_{11}$ emission transition (985 nm)[20]. Figure 3a shows bright individual (6,5) nanotubes displaying uniform luminescence.

The selected nanotubes are then imaged using the modulated absorption method. Since the signal amplitude and profile depend on the nanotube orientation with respect to the beam modulation axis, the measurements are performed only on long nanotubes perpendicular to y-axis (Figure, 3b). Furthermore, the laser polarization is set along the nanotube for maximum interaction[21]. To ensure that the nanotubes are isolated from other nanotube species or from catalyst impurities, always abundant in nanotubes samples, we performed photothermal heterodyne microscopy on the same sample region[22, 23]. In this case, we use circularly polarized excitation to obtain images independent from tube orientations (Figure 2c and Figure 3S, Supp. Info). The perfect correlation between PL, photothermal and modulated absorption images allows assigning the individual (6,5) nanotubes without ambiguities.

In contrast to gold nanoparticles, which can be considered infinitely small compared to the beam profile, the 1D geometry of the nanotube has to be taken into account in the derivation of demodulated power: $P_{mod}(x,y) = -\sqrt{\frac{8}{\pi}} \frac{\delta}{w_0^3} P_{inc} n_C \sigma_C y e^{\frac{-2y^2}{w_0^2}} \left[ erf\left[\frac{\sqrt{2}}{w_0}(x+\frac{L}{2})\right] - erf\left[\frac{\sqrt{2}}{w_0}(x-\frac{L}{2})\right]\right]$

where $L$ is the SWCNT length and $n_C$ the number of carbon atoms per tube length ($n_C$ = 88271 μm$^{-1}$ for (6,5) nanotubes). Figure 3d displays the signal profile $P_{mod}(x=0,y)$ along the y-axis measured around the middle part of the tube at excitation power of 70 μW.

In the case of nanotubes significantly longer than the diffraction limit (i.e. L>1μm), $\sigma_C$ can be expressed as a function of $A_{pp}$ using: $\sigma_c = \frac{\sqrt{2\pi}e^{1/2}}{8} \frac{w_0^2}{n_c \delta} \frac{A_{pp}}{P_{inc}}$. The excitation laser wavelength is being not strictly resonant with the S$_{22}$ transition; one has to account for the dependence of the absorption on the excitation wavelength to provide a value for $\sigma_C$ at the peak of this transition. For this purpose, we acquired photoluminescence excitation (PLE) of individual (6,5) nanotubes prepared following the same sample procedure as for absorption measurements because the optical spectra depend on the nanotube local environment. A tunable dye laser (emission range 540-590 nm) is used for excitation and the SWCNT luminescence signal was recorded at low excitation intensities to avoid any saturation effect. PLE spectra displayed systematically a Lorentzian profile[24] with a peak S$_{22}$ transition found at 569±2 nm (Figure 4S, Supp. Info). From 8 spectra acquired, a correction factor of 1.47 ± 0.12 is deduced allowing the determination of $\sigma_C$ at the S$_{22}$ peak transition using the measurements performed with low noise solid-state laser at 561 nm. Figure 3e displays the histogram $\sigma_C$ measured on 73 different (6,5) nanotubes. A unimodal distribution is obtained providing a clear signature that all data points stem from individual chirality identified nanotubes. The histogram is well adjusted with a Gaussian distribution centered at 3.2 10$^{-17}$ cm$^2$ with a half-width-at-half maximum of 0.5 10$^{-17}$ cm$^2$. This value compares to the upper range of previous indirect estimations of $\sigma_C$ (Figure 3f)[14]. Importantly, this obtained value originates from direct measurements (model independent), therefore, we can determine both its precision and its exactitude in a straightforward manner. The relative error on the predetermination of $w_0$ (2%, which translate to 4% error on $\sigma_C$), $\delta$ (5%) and the excitation wavelength correction factor (9%) might induce a systematic bias on the determination of $\sigma_C$ thus affecting its exactitude by up to 18%. Furthermore, we estimate the precision of this measured value to be 15% stemming from as a 10% noise-to-signal ratio for $A_{pp}$, a 1% precision in the determination of the beam power and 2% fluctuations on $w_0$ due to sample-to-sample focusing. We thus conclude that the measured spread of the data (±16 %) displayed on figure 4b reflects the precision on the determination of $\sigma_C$ while its exactitude lies within 18%.

Correlative imaging (luminescence versus absorption) of long carbon nanotube provide a direct evidence of exciton diffusion and a straightforward determination of the exciton diffusion length before its recombination[25, 26]. Previous determinations rely on stepwise quenching of nanotube luminescence by chemical reactions[27, 28], by ensemble studies of the length dependence of PL efficiency[29] or on studying the luminescence spatial profile at nanotube ends[30]. In the latter case, determination of $l_D$ relies on fitting carefully the intensity profile at the nanotube end where local quenching occurs.

Here we show that by simultaneous acquisition of the luminescence and absorption profiles, $l_D$ can be directly determined by confronting the apparent nanotube ends between the two imaging modalities. To this aim, individual nanotubes were imaged by confocal luminescence and spatial modulation absorption microscopies with the same confocal excitation beam (Figure 4a-b). Comparison of the images reveals that nanotube PL profiles appear systematically shorter than absorption ones (Figure 4c). This is a direct consequence of exciton diffusion. The apparent nanotube ends appear separated by ~ 150 nm between the two images. Using a simple 1D exciton diffusion model (Supp. Info) with quenching at nanotube ends, we simulated the expected PL and absorption profiles obtained in our confocal microscope for $l_D$ varying from 0 to 500 nm. The best agreement with the experiments is found for $l_D \approx 200$ nm (figure 4d). This value is in agreement with previous determinations for nanotube suspended in DOC[30, 31].

In conclusion, we measured the absorption cross-section of (6,5) carbon nanotubes at their second order optical transition and obtained a value of 3.2 $10^{-17}$ cm$^2$ per carbon atom. This constitutes the first metrological investigation of the absorption cross-section of chirality-identified nanotubes. Beside its importance for fundamental investigations of nanotube physical properties (e.g. accurate determination of exciton rate generation), this value is also essential for carbon nanotube material science and a key parameter for exploiting the large application potential that nanotubes offer in optoelectronics and photovoltaics.

ACKNOWLEDGMENT. This work was funded by the Agence Nationale de la Recherche, Région Aquitaine, DGA and the European Research Council.

§ present address: Los Alamos National Laboratory, Los Alamos, NM 87545

FIGURES CAPTIONS

Figure 1: Schematics of the experimental setup. A sample placed at the focal plane of a microscope objective and containing isolated nano-objects (gold nanoparticle or carbon nanotubes) is mounted on a piezo-scanner. An acousto-optic modulator is used to modulate the position of the exciting beam in the sample plane (beam modulation amplitude of 36nm along the y-axis at 100kHz). The transmitted beam is collected by second microscope objective and sent to a fast photodiode connected to a lock-in amplifier. The signal is the demodulated power $P_{mod}$ as a function of the sample position. A typical profile shape of $P_{mod}$ along the y-axis is illustrated (see text).

Figure 2: a) TEM image of 10nm individual gold NP used in this study. b) Histogram of the diameters measured on 191 NP and its Gaussian fit. c) Modulated absorption image of a single 10 nm gold NP excited at 561 nm ($P_0$ = 650 µW, 25ms/pixel). d) Corresponding profile averaged over 4 lines of c). e) Histogram of the absorption cross-sections deduced for 100 individual 10 nm gold NPs with its Gaussian fit (see text).

Figure 3: Luminescence (a, 300 ms integration time), photothermal (b, 5 ms/pixel) and modulated absorption (c, 25 ms/pixel) images of the same individual (6,5) carbon nanotube. Scale bars 1µm. d) Absorption profile perpendicular to the nanotube axis averaged over 10 lines of c). e) Histogram of the absorption cross sections deduced for 71 individual (6,5) carbon nanotubes and its Gaussian fit (see text). The mean value of the distribution is 3.2 $10^{-17}$ cm$^2$ and its standard deviation 0.5 $10^{-17}$ cm$^2$. f) Comparison with the values reported in the literature for (6,5) nanotubes (all values were normalized to an excitation at the $S_{22}$ transition with a polarization along the nanotube axis): △[6]; ▽[7]; □[8]; ◇[11]; ○[9]; ☆[14]; ● this work.

Figure 4: Confocal microscopy images ((a) modulated absorption, (b) luminescence, 50ms/pixel) of an individual (6,5) carbon nanotube using the same excitation beam. c) Corresponding absorption (black) and luminescence (red) profiles along the nanotube axis. d) Simulated profiles using a 1D diffusion equation (see text and Supp. info): absorption (black) and photoluminescence (red and blue) for varying diffusion lengths (100 to 500 nm). The best agreement with experiments is obtained for $l_D$ =200 nm (red).

Figure 1

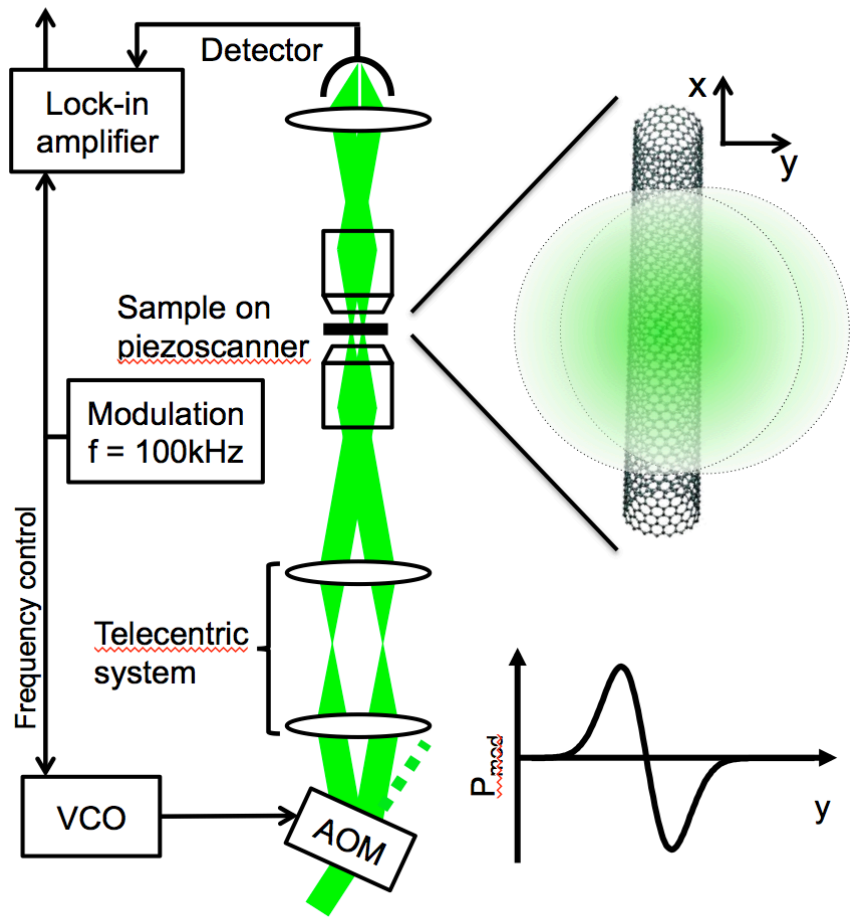

Figure 2

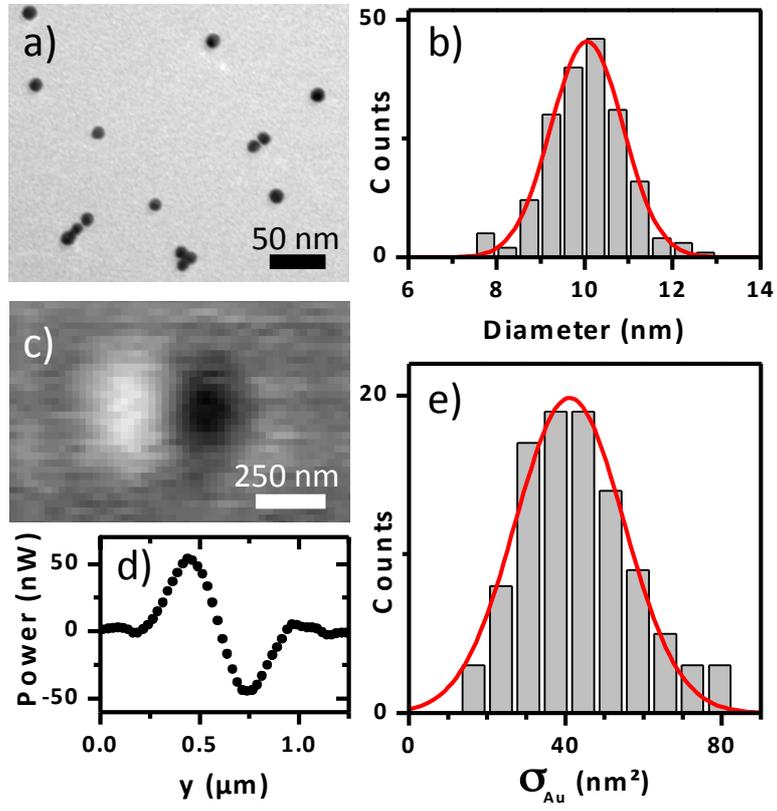

Figure 3

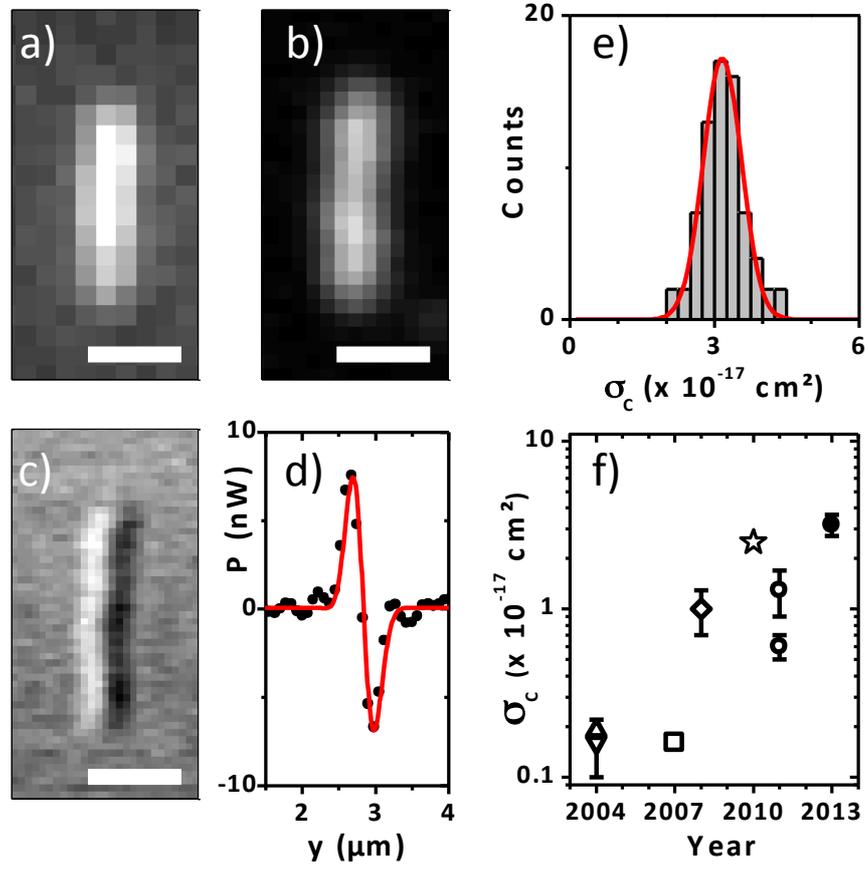

Figure 4

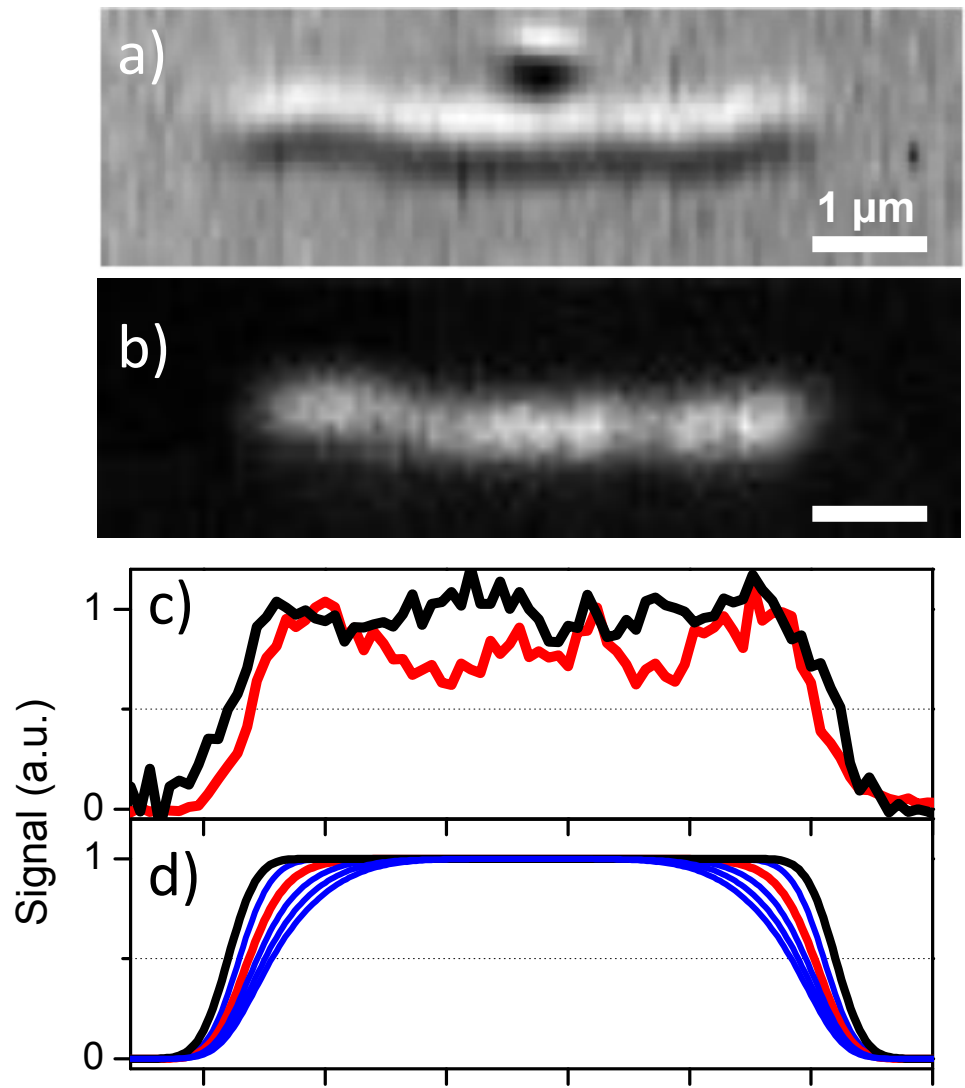